\documentclass[aps,prl
]{revtex4}
\def\frac#1#2{{\textstyle{#1\over#2}}} 

\def\ket#1{| #1\rangle}

\def\R{\hbox{\rm I \kern-5pt R}}

\usepackage{amssymb}
\usepackage{amsmath}
\usepackage[sort&compress]{natbib}
\usepackage[dvips]{graphicx}

\begin{document}


\title{Quantum Tagging: Authenticating Location via Quantum Information and Relativistic
Signalling Constraints}
\author{ Adrian Kent}
\affiliation{Centre for Quantum Information and Foundations, DAMTP, University of Cambridge, 
Cambridge, U.K.}
\affiliation{Perimeter Institute for Theoretical Physics, Waterloo, Ontario, Canada}
\author{William J. Munro}
\affiliation{NTT Basic Research Laboratories, NTT Corporation, 3-1 Morinosato-Wakamiya, 
Atsugi-shi, Kanagawa 243-0198, Japan}
\author{Timothy P. Spiller}
\affiliation{Quantum Information Science, School of Physics and Astronomy, University of Leeds, Leeds LS2 9JT, U.K. }
\date{ August 2010; published version with updated references and typo fixes July 2011 }

\begin{abstract}
We define the task of {\it quantum tagging}, that is, authenticating the classical location of a 
classical tagging device by sending and receiving quantum signals from suitably located distant sites,
in an environment controlled by an adversary whose quantum information processing and transmitting power
is unbounded.  We define simple 
security models for this task and briefly discuss alternatives. 

We illustrate the pitfalls of naive quantum cryptographic
reasoning in this context by describing several protocols 
which at first sight appear unconditionally secure but which, as we show,
can in fact be broken by teleportation-based 
attacks.   We also describe some protocols which cannot be broken by
these specific attacks, but do not prove they are
unconditionally secure.  

We  review the history of quantum tagging protocols, which we 
first discussed in 2002 and described in a 2006 patent (for an insecure
protocol).  The possibility has recently been reconsidered by Malaney and Chandran et al. 
All the more recently discussed protocols of which we are aware were either previously
considered by us in 2002-3 or are variants of schemes then considered, and all are provably insecure.  

\end{abstract}
\maketitle
\section{Introduction} 

There is now a great deal of theoretical and practical interest in the
possibility of basing unconditionally secure cryptographic tasks on 
some form of no-signalling principle as well as, or even instead of, the laws
of non-relativistic quantum theory.   The earliest examples of which we are
aware are 
bit commitment protocols based
on no-signalling \cite{kentrel, kentrelfinite}, discovered in 1999-2000, 
which are provably secure against all classical attacks
and against Mayers-Lo-Chau quantum attacks.   
The first secure quantum key distribution
protocol based on no-signalling\cite{bhk} was discovered in 2005. (See also Ref.~\cite{BKP} for some further
  details and discussion.) It 
has subsequently been significantly developed,
producing more efficient protocols provably secure against restricted
classes of attack~\cite{AGM,AMP,ABGMPS,PABGMS,McKague} and then
against general
attacks~\cite{MRC,Masanes,HRW2,MPA,HR}.  A protocol for an interesting novel cryptographic
task, variable bias coin tossing, using both quantum theory and the 
no-signalling principle, was published in 2006 \cite{vbct}.   
Protocols for expanding a private random string, using untrusted
devices, based on quantum theory and the no-signalling principle, were
also recently introduced (\cite{ColbeckThesis}; see \cite{ckrandom}
for a more complete presentation of this work) and
significantly developed\cite{PAMBMMOHLMM,ckrandom}: note that at present
the unconditional security 
of all these randomness expansion protocols against completely general attacks is an open 
question.  

We define and discuss here another example of an interesting cryptographic task,
quantum tagging, and present and discuss quantum tagging protocols which 
rely both on 
the properties of quantum information and on the impossibility of 
superluminal signalling.   

\section{Quantum tagging: definitions} 

We work within a Minkowski space-time $M^{(n,1)}$, with $n$ space dimensions.
The most generally applicable case is thus $n=3$.\footnote{Corrections due to 
general relativity in weak gravitational fields can be made without qualitatively 
affecting our discussion \cite{kentrelfinite}.  In particular, our discussion
applies to tagging protocols on or around Earth, assuming that it is reasonable
to neglect exotic gravitational phenomena 
such as wormholes or attacks based on major distortions of the local space-time geometry.}
The case $n=2$ applies in scenarios where all parties are effectively restricted to a plane -- for instance
a small region of the Earth's surface. 
The case $n=1$ is physically realistic only if the agents are effectively 
confined to a line.  Although this is unlikely in realistic applications, 
we will consider this case below,
as it simplifies the discussion while
illustrating many key points.  

There are several distinct interesting security scenarios for
quantum tagging, for example: 

\subsection{Security scenario I}

Alice operates cryptographically secure sending and receiving
stations $A_0$ and $A_1$, located in small
regions (whose size we will assume here is negligible, to simplify
the discussion)
around distinct points $a_0$ and $a_1$ on the real line. 
The locations of these stations are known to and trusted by Alice,
and the stations contain synchronized clocks trusted by Alice.
Her tagging device $T$ occupies a finite region $ [ t_0 , t_1 ]$
of the line in between these stations, so that $a_0 < t_0 < t_1 < a_1 $.   The tagging device
contains trusted classical and/or quantum receivers, computers and
transmitters, which are located in a small region (which again, to
simplify the discussion, we assume is 
of negligible size compared with $(t_1 - t_0)$ and other parameters)
around the fixed point $t_+ = \frac{1}{2} ( t_0 + t_1 )$.
The device is designed to follow a protocol in which classical
and/or quantum outputs are generated via the computer from inputs defined
by the received signals.   The outputs are sent in a direction, left
or right (i.e. towards $a_0$ or $a_1$) that again depends on the inputs.
The tagging device may also contain a trusted clock,
in which case the clock time is another allowed input.\footnote{
One simple use for a trusted clock within $T$ is to verify that 
the correct number of signals is received within a given time interval.
This prevents $E$ from generating, and causing $T$ to process,
more signals than allowed by the protocol.  Although we do not 
immediately see scope for $E$ to obtain significant extra power
by exploiting this option in the protocols
considered below, excluding it would, at the very least, simplify the 
security analysis.}
Note however that it would not make sense in our scenario to {\it assume} from
the start 
that the tagging device $T$ contains a trusted GPS device so that $T$ can 
verify and authenticate its own location. 
To analyse possibilities of this type, we would need to include the fixed
GPS stations among Alice's laboratories, and the communications between
these stations and $T$ would form part of the tagging protocol.

We assume that signals can be sent from $A_i$ to $T$, and within $T$, at
light speed, and that the time for information processing within $T$ 
(or elsewhere) is negligible.   $T$ is assumed {\it immobile} and {\it physically secure}, in the sense that an
adversary Eve can neither move it nor alter its interior structure.
However, $T$ is not assumed {\it impenetrable}: Eve may be able to 
send signals through it at light speed, and may also be able to inspect its
interior.   In particular, $T$ contains no classical or quantum
data which Alice can safely assume secret, and she must thus assume
that its protocol for generating outputs from inputs is potentially
public knowledge.

$T$ can be switched on or off.   When switched off, it remains immobile
and physically secure, and simply allows any signals
sent towards it to propagate unmodified through it: in particular, signals
travelling at light speed outside $T$ also travel through $T$ at light speed. 

Eve may control any region of space outside $A_i$ and $T$, may send classical
or quantum signals at light speed through $A_i$ and $T$ without $A$ (or $T$) detecting them, may
be able to jam any signals sent by $A_i$ or $T$, and may carry out arbitrary 
classical and quantum operations, with negligible computing time, anywhere
in the regions she
controls.   Eve cannot cause any information processing to take place within $T$,
other than the (computationally trivial) operation of transmitting arbitrary signals through
$T$, except for the operations that $T$ is designed to carry out on appropriate input signals.   
Her task is to find a strategy which {\it spoofs} the actions of $T$, that is, 
makes it appear to $A$ that $T$ is switched on when it is in fact
switched off.    Conversely, $A$'s task is to design $T$, together with a tagging
protocol with security parameter $N$, so that the chance, $p(N)$, of 
$E$ successfully spoofing $T$ throughout a given time interval $\Delta t$
obeys $p(N) \rightarrow 0$ as $N \rightarrow \infty$.   

In this scenario, Eve is limited: she can neither move $T$ nor carry out non-trivial
operations within the space it occupies.  One could imagine that $T$ is tagging an
object in a hostile environment which neither $E$ nor $A$ can enter.  
$E$ might, however, be able to destroy the object together with $T$ --
thus effectively switching $T$ off -- and spoof the tagging protocol so that
$A$ is unaware of the loss.   

\subsection{Security scenario II}

In scenario II, the tag is physically secure, but not immobile.  
Eve can move it, without disturbing its inner workings, at any
speed up to some bound $v$, known to Alice.   
Clearly $v=c$, the speed of light, gives an absolute upper
bound.   To avoid considering relativistic effects, 
we assume $v \ll c$ here when we consider this 
scenario.\footnote{
In practical applications, if it is reasonable to assume
the physical security of $T$ --- i.e. if this security model
is relevant at all --- we believe it will not be hard to 
establish and justify such a bound for a protocol of finite
time duration.   For example, a bound on the initial
velocity of $T$, together with a bound on the possible
acceleration that $E$ can impart while maintaining the 
integrity of $T$, thus imply a bound for $v$ at any given
later time.}

\subsection{Practical relevance}  

As already noted, in realistic applications, $T$ would generally 
occupy a $3$-dimensional region, $A$ might have any number of 
sending and receiving stations lying in different directions from
$T$, and $T$'s outputs might be sent to any or all of these.   

We envisage that in realistic applications $T$ would be a device securely attached to 
an object whose location is significant to $A$.   In practice, we imagine,
Eve might be able to destroy $T$, or move it along with the object to a region disjoint
from that it originally occupies, and then replace it with another device. 
However, each of these operations would necessarily take some time, and 
we assume the relevant time can be bounded below by some minimum, $\Delta t$.\footnote{
For example, the time needed to move $T$ to a region disjoint from its original
location is at least $L/c$, where $L$ is the minimum diameter of $T$.}
The idea of a tagging protocol 
is thus to ensure that any such interference by Eve would be  
detected by Alice before Eve's operations are complete, because $T$ is not 
functioning as it should, according to the protocol, given its 
presumed location.   Within a given security scenario, tagging protocols in which
$A$ is attempting to verify that $T$ is stationary 
can easily be generalised to protocols in which $A$ is attempting to
verify the location of $T$, when she knows that $T$'s speed will be bounded
(with respect to a given frame, for example
the stationary frame of Alice's stations).  We thus
consider the case of verifying the location of a stationary $T$.  

In this way, we separate the issues of $T$'s physical security and the 
security of its attachment to the object from specific aspects of
its cryptographic security, defined by appropriate models.   
We analyse cryptographic security here via the security models 
given above.\footnote{
Incidentally, quite apart from their cryptographic applications,
these models seem to us to raise interesting new questions
about the properties of quantum theory and quantum 
information in Minkowski space.}
We would argue that, in a scenario in which all a tag's operations are potentially
visible to an adversary, a tagging protocol which is provably breakable in
one of our models cannot be sensibly said to be unconditionally secure.
To analyse the physical and other security issues in realistic applications,
one needs further to consider how well -- and under which assumptions --
these models apply.   We do not examine these latter issues
further here (but see Ref. \cite{kentcryptotagging} for further discussion).  

Other security scenarios and other models can also be considered, of course.   
Our aim here is to introduce the problem of quantum tagging and set out
some interesting scenarios and questions, not to analyse all possibilities.   

\subsection{Spoofing}

In a general spoofing attack on a tagging scheme, Eve intercepts
some or all of the signals transmitted by $A$ and $T$ at one or more 
sites, carries out information processing on them at these sites,
and retransmits the resulting outputs, which may be rerouted or delayed,
to other sites under her control and/or to $A$ and/or $T$. 
Her information processing may involve collective operations on any
information in her possession, including signals received directly from $A$ and $T$,
ancillary information generated in her sites, 
and information generated by her own earlier operations.  

For example, tagging schemes that do not rely on precise timings
are vulnerable to simple 
{\it record-and-replay} spoofing attacks.
In a record-and-replay attack, Eve intercepts all the
outgoing signals from the tagging device, in a way that
effectively jams the outgoing channel, preventing any 
signal reaching $A$ from $T$.  Eve then  
replays the outgoing signals, unaltered, at  
later times, transmitted from different locations.  
By so doing she can hope to persuade $A$ that the 
device is in a given location when its location
has in fact been altered: i.e., she can hope 
to render the scheme insecure under scenario II above.  

Our aim is to discuss the possibility of devising protocols that use timed 
quantum (and perhaps classical) signals, together with 
relativistic signalling constraints, to ensure security 
against general spoofing attacks.   

\subsection{Types of input and output}

We want to distinguish between input and output signals
that carry classical information and those that 
carry quantum information.  By the latter, we mean
signals carried by a single quantum state lying in
a fixed finite-dimensional Hilbert space --- for
example, a qubit.  By the former, we mean a 
signal robust enough and redundant enough to be
considered classical, that can be copied effectively infinitely
and broadcast with effectively arbitrary fidelity, and 
that cannot practically be created in superposition: 
for example, a radio transmission.  

Physics (as currently understood) provides no fundamental qualitative distinction
between the classical and quantum.  Any classical signal
could be treated as a (perhaps very redundant) quantum 
signal, by considering a Hilbert space of suitably 
large dimension.  Nonetheless it would be practically
significant and cryptographically interesting to find a scheme that
is secure if (but only if) some signals are considered classical. 

To simplify the analysis a little, we characterise 
an input which involves both 
type of signals --- for example, a
classical input from one source and a quantum input
from another --- as a {\it quantum input}, 
and we characterise a {\it quantum output} 
similarly.  This gives four distinct cases to consider:
classical input and classical output (CC), quantum
input and classical output (QC), classical input
and quantum output (CQ), and quantum input
and quantum output (QQ).  
Since CC and CQ schemes allow the input to be copied and
broadcast, creating an immediate potential vulnerability, we focus here on QC and
QQ schemes.   

\section{Some simple insecure schemes}

The schemes we describe in this section are 
not perfectly secure.  We nonetheless find them
of practical and theoretical interest,
since the only attacks to which we know they
are vulnerable require advanced information technology that is presently
unavailable (specifically, perfectly
efficient implementation of quantum teleportation).  
We assume noiseless communication here:
our discussion can be generalized to the noisy case
by considering standard error correction methods.  

\subsection{Scheme I}

Alice sends quantum
signals, taking the form of a series of independently randomly chosen
qubits, $\ket{\psi_i}$, from $A_0$, and classical signals, taking the form of
a series of independently randomly chosen bits, $a_i$, from $A_1$.
The qubits are chosen to be pure states, drawn randomly from the
uniform distribution on the Bloch sphere. 
These signals are sent at light speed, timed so as to arrive pairwise
simultaneously at $t_+$: that is, the first qubit and the first
bit arrive together, then the second qubit and the second bit,
and so on. 

The tagging device $T$ interprets the classical bits as 
an instruction to send the qubit $\ket{\psi_i}$ in the direction
of $A_0$ or $A_1$ (i.e. $a_i$ codes to send towards
$A_{a_i}$).  Upon receiving the bit and qubit, $T$ immediately
obeys the instruction, redirecting the qubit in the 
appropriate direction.   Alice tests that the qubits
received at the receivers $A_i$ are the qubits she
sent, and that they arrived at the appropriate times.
(The first test is implemented by carrying out a projective
measurement onto the space spanned by the originally
transmitted qubit.)   If this test is passed for $N$ successive
qubits, sent within the interval $\Delta t$, 
she accepts the location of $T$ as authenticated. 

\subsection{Scheme II}

Alice sends a sequence of pairs 
$ (a_i  , \ket{ \psi_i } ) $ from $A_0$, and a 
sequence $ b_i $ from $A_1$.  Here the $a_i$ are
a sequence of independently randomly chosen 
numbers in the range $ 1 \leq a_i \leq m$, and
the $b_i$ are
a sequence of independently randomly chosen 
numbers in the range $ 1 \leq b_i \leq n$,
while the $\ket{\psi_i }$ are independently
randomly chosen qubits.   
The qubits are chosen to be pure states, drawn randomly from the
uniform distribution on the Bloch sphere. 

The signals $ (a_i , \ket{\psi_i } )$ and $b_i$ 
are timed to arrive pairwise simultaneously at $t_+$. 
The $a_i$ and $b_i$ together code an instruction,
defined by some previously fixed function
$f( a_i , b_i ) \in \{ 0, 1 \}$, 
to send the qubit to detector $A_0$ or $A_1$ respectively.\footnote{If we consider 
more than one spatial direction, 
this scheme can be generalised to allow the qubit to be sent
to any of a number of laboratories lying in different directions from $T$.} 
Immediately on receipt of the $i$-th set of
signals, $T$ follows this instruction, 
redirecting the qubit towards $A_{f (a_i , b_i )}$.   Alice tests that the qubits
received at $A_i$ are the qubits she
sent, and that they arrived at the appropriate times.
(The first test is implemented by carrying out a projective
measurement onto the space spanned by the originally
transmitted qubit.)   If this test is passed for $N$ successive
qubits, sent within the interval $\Delta t$, she accepts the location of $T$ as authenticated. 

\subsection{Scheme III}

Alice sends a sequence of independently
randomly generated qubits $\ket{\psi_i}$ from $A_0$, and a sequence
of independently randomly generated 
classical trits $c_i$ from $A_1$. 
The qubits $\ket{\psi_i}$ are chosen randomly from
the set $\{ \ket{0}, \ket{1}, \ket{\pm}, \ket{\pm i} \}$,
where 
$$
\ket{\pm} = \frac{1}{\sqrt{2}} ( \ket{0} \pm \ket{1} ) \, , 
\ket{\pm i} = \frac{1}{\sqrt{2}} ( \ket{0} \pm i \ket{1} ) \, . 
$$
The trits, which are uniformly distributed, are interpreted as coded instructions
to carry out a projective measurement in one of
the bases $ B_0 = ( \ket{0}, \ket{1} )$, 
$B_1 = (\ket{+} , \ket{-} )$, $B_2 = ( \ket{i}, \ket{-i} )$. 
These signals are sent at light speed, timed so as to arrive pairwise
simultaneously at $t_+$: that is, the first qubit and the first
bit arrive together, then the second qubit and the second bit,
and so on. 

As soon as the pair $\ket{\psi_i}$ and $c_i$ are received,
$T$ measures $\ket{\psi_i}$ in 
the basis $B_{c_i}$.  It then immediately classically broadcasts 
the measurement outcome bidirectionally.\footnote{In more than one spatial
dimension, $T$ would broadcast the outcome omnidirectionally.}

If the measurement statistics agree with those predicted
by quantum theory, and the measurement results are received
at the appropriate time by both detectors, for $N$ successive
qubits, sent within the interval $\Delta t$, Alice accepts the location of $T$ as authenticated. 

\subsection{Comment on authentication and timing}

For these schemes, and indeed any quantum tagging scheme in which Alice maintains
laboratories at separated sites, it will of course take some time for her to collate and
compare the data received at her various sites.   Alice thus cannot possibly hope to
authenticate, at any given time, that the tagging device is functioning correctly {\it at
that instant in time} (in her lab rest frame).   The aim of a quantum tagging protocol
is rather to allow her to verify that the device was functioning correctly, at the 
correct location, within a given past fixed time interval (which necessarily lies in the
past light cone of the point(s) at which verification is completed).   

\subsection{Discussion of (in)security of schemes I-III}

An argument, which may seem plausible at first sight, suggests that 
schemes I-III should be unconditionally secure.
We first review the argument, then explain why it is incorrect
and show that the schemes are in fact insecure in either
of our security models.   

\subsection{A naive security argument}

Because quantum information cannot be cloned, the
incoming qubits $\ket{\psi}$ must follow a unique
path.  If the qubits are transmitted directly to 
$T$, the required output data cannot be reliably
generated at the correct time except by following 
the tagging protocol at $T$.   If, on the other
hand, the qubit is rerouted in some other direction, 
it encounters the classical data transmitted from $A_1$ 
at a later time than it would have if the tagging
protocol were followed.  

Since Eve does
not know what to do with the qubit -- which way to
send it, or in which basis to measure it -- until
the classical data from $A_1$ arrive, she
cannot be sure how to act until the classical and
quantum data coincide.  By this point, it is too
late for them to be able to produce outputs that
will arrive at the correct times at both $A_0$ and $A_1$. 
For example, in scheme I, if $E$ delays and
stores the qubit somewhere between $A_0$ and $T$, and
waits for the classical signal to arrive from $A_1$ before
retransmitting the qubit, she can spoof the protocol if instructed
to send the qubit to $A_0$, but cannot get the
qubit to $A_1$ in time if instructed to send the qubit there.

Hence -- the argument purports to show -- on each round 
any spoofing attack has a nonzero probability of detection. 
Moreover, a nonzero lower bound for this probability can be calculated,
and the tagging scheme is thus secure. 

\subsection{What's wrong with the naive security argument?}

One could try to formalise this naive argument
as follows.  Because of the no-cloning theorem,
the quantum information encoded in the qubits
$\ket{\psi_i}$ cannot be duplicated and so
must follow a single definite trajectory.
This would imply in particular that the quantum information
must be localised at a single point at any given time.  

While this might seem plausible at first sight, 
there are actually many ways in which quantum information
can be delocalized.  For example, $E$ could create a superposition
of distinct trajectories via interferometry. 
Another possibility is that she could teleport the qubit and 
broadcast the classical information generated by the teleportation.  
These possibilities show that the naive security argument fails, 
since after these operations the quantum information encoded
in the qubit no longer follows a single space-time path, at least in 
any standard sense.  We now show that
$E$ can indeed exploit the power of teleportation to break the above
schemes.  

\subsection{Teleportation attacks on schemes I and II}

As scheme I is a special case of scheme II, we need
only consider the latter.
Consider the following attack.  

Eve sets up laboratories at sites
$E_0$, between $A_0$ and $T$, and $E_1$, 
between $T$ and $A_1$.  
She arranges a sequence of labelled entangled
singlet pairs to be shared between the sites
$E_0$ and $E_1$, with labels $i$ (indicating which tagging
signal a given set of pairs is going to
be used to attack) and $j$ (which runs from
$1$ to $m$).   When the signal $( a_i , \ket{\psi_i})$
reaches $E_0$, Eve carries out a teleportation measurement with the incoming qubit and the 
first singlet qubit with label $(i,a_i)$. The classical teleportation data, describing the 
unitary operation needed to complete the teleportation, are immediately sent towards $E_1$ with a copy 
being kept at $E_0$.\footnote{In higher dimensions, the classical teleportation data would be 
broadcast to suitably located stations controlled by Eve.}
When the signal $b_i$ reaches $E_1$, Eve 
sends all the qubits stored there with labels $(i, a_i )$ 
for which $f(a_i , b_i ) = 0$ towards the site $A_{0 }$, using distinct physical
degrees of freedom so that she can identify each qubit's
label $a_i$ at any later time.   She stores at $E_1$ all the qubits with labels $(i, a_i )$ 
for which $f(a_i , b_i ) = 1$, until receipt of the signal $a_i$.
When the signal $a_i$ and the teleportation signal simultaneously reach $E_1$, if $f(a_i , b_i ) =1 $, 
then Eve (at $E_1$) applies the teleportation operation to the stored qubit with label $(i, a_i)$ 
and transmits the teleported qubit towards $A_1$, discarding the remaining qubits from batch $i$ stored
at $E_1$; 
if $f(a_i , b_i ) = 0$ then Eve (at $E_1$) discards all the qubits from batch $i$ stored at $E_1$. 
When the signal $b_i$ and the transmitted qubits simultaneously reach $E_0$, if $f(a_i , b_i ) =0$, then
Eve (at $E_0$) applies the teleportation operation to the transmitted qubit with label $(i, a_i )$,
transmits this qubit towards $A_0$, and (at $E_0$) discards the others from batch $i$ stored at $E_0$; if $f(a_i , b_i ) =1$ she
discards all the qubits from batch $i$ stored at $E_0$.  

Eve attempts to ensure that none of her classical signals are detected by $A_0$ or $A_1$, 
either by transmitting them on frequencies not used by $A$ or by jamming her classical signals
so that none is transmitted to the left of $E_0$ or the right of $E_1$.\footnote{In practice, 
whether Eve succeeds in this may depend on the technology available to her and Alice.   
However, for the protocol to be unconditionally secure, we require that an Alice whose signal detection power is bounded
must be able to detect a spoofing attack by an Eve whose signalling technology is
unbounded.}
Eve allows Alice's classical signals to propagate freely between $A_0$ and $A_1$ -- i.e. she
reads them but does not jam them.    

Through this teleportation attack, Eve can spoof the tagging scheme and cause Alice to accept the location of $T$ 
as authenticated.

\subsection{Teleportation attacks on scheme III}

Eve sets up laboratories at sites
$E_0$ and $E_1$, located as above.  
She arranges a sequence of labelled entangled
singlet pairs to be shared between these sites, 
with labels $i$ (indicating which tagging
signal a given set of pairs is going to
be used to attack). 
When the signal $\ket{\psi_i}$ arrives at $E_0$ from
$A_0$, she carries out a teleportation measurement
at $E_0$.  
The classical teleportation data, describing
the unitary operation needed to complete the
teleportation, are immediately sent towards
$E_1$, with a copy being kept at $E_0$.  
When the signal $c_i$ arrives at $E_1$
from $A_1$, she carries out a measurement in basis
$B_{c_i}$ on the second particle from singlet $i$. 
The measurement outcome
and basis are immediately sent towards $E_0$. 

The teleportation
unitary operations $I, X, Z, XZ$
leave the bases $B_0$, $B_1$
and $B_2$ invariant.\footnote{They may map one
basis state to the other, but the unordered
set containing the pair of states is left unchanged.}
Hence, from the outcome of a measurement in a basis
$B_j$ ($j = 0,1$ or $2$) on the
unitarily rotated state $U \ket{\psi_i}$ represented
by the second entangled qubit, together with a 
description of the unitary $U$, Eve 
can infer the outcome of the same measurement  
on the original state $\ket{\psi_i}$.  

Thus, combining the classical signals from $E_0$ and
$E_1$ at either site, Eve can infer the measurement outcomes 
required by the tagging scheme.
By sending these outcomes immediately to the $A_i$, 
she can thus spoof the tagging scheme. 
 
\section{Secure tagging protocols?}

The vulnerability of schemes I and II
to the teleportation attacks described reflects a general weakness
of QQ schemes in which the output is directed to a
single detector.  The attack described on 
scheme III reflects a specific weakness in the
design of this scheme, arising from the fact that the 
measurement bases chosen are invariant under teleportation
operations.  This motivates considering variations
on this scheme, such as the following examples.  

\subsection{Scheme IV}

From $A_0$, Alice sends a sequence of random
pure qubit states, drawn independently from the
uniform distribution on the Bloch sphere.
From $A_1$, she sends a classical signal selecting
a random measurement basis, drawn independently
from the uniform distribution on the set of 
Bloch sphere antipodes (i.e. uniformly distributed on some
hemisphere).  These states are sent to arrive simultaneously
at $T$, which is instructed to 
carry out a measurement of the received qubit
in the specified basis and then 
immediately to broadcast the
outcome classically to both $A_0$ and $A_1$. 

\subsection{Scheme V}

Clearly, scheme IV is idealized: a real implementation
would select the qubit and basis from (perhaps very
large) finite lists, approximating uniform distributions
over the Bloch sphere.   This can be done in infinitely
many ways.   Scheme V is one concrete and simple example,
using a simplified version of scheme IV.  It exploits
the essential idea, without attempting a good approximation
of uniform distributions.  
From $A_0$, Alice sends random states drawn
from the list $\{ \, \ket{0}, \ket{1}, 
\cos ( \pi/6 ) \ket{0} + \sin (\pi /6 ) \ket{1} , 
\sin ( \pi/6 ) \ket{0} - \cos (\pi /6 ) \ket{1} , 
\cos ( \pi/6 ) \ket{0} + i \sin (\pi /6 ) \ket{1}, 
\sin ( \pi/6 ) \ket{0} - i \cos (\pi /6 ) \ket{1} \, \}$.  

From $A_1$, she sends random trits coding for
measurements in the bases 
$B'_0 = \{ \ket{0}, \ket{1} \}$, 
$B'_1 = \{ \cos ( \pi/6 ) \ket{0} + \sin (\pi /6 ) \ket{1} , 
\sin ( \pi/6 ) \ket{0} - \cos (\pi /6 ) \ket{1} \}$, 
$B'_2 = \{ \cos ( \pi/6 ) \ket{0} + i \sin (\pi /6 ) \ket{1} , 
\sin ( \pi/6 ) \ket{0} - i \cos (\pi /6 ) \ket{1} \}$.  
The scheme then proceeds as above.

\subsection{Scheme VI}

Scheme VI is a variation of
scheme IV, with an extra feature which may  
make the security of the scheme easier to prove
(if indeed it is provably secure).   
The same idea can be used to define variations
of scheme V or other schemes related to scheme
IV.  

From $A_0$, Alice
sends random states $\ket{\psi_i}$
drawn independently from the uniform distribution on the
Bloch sphere.  The classical signal broadcast from $A_1$ sends
a random measurement basis $b$, drawn independently
from the uniform distribution on the set of 
Bloch sphere antipodes (i.e. uniformly distributed on some
hemisphere) and two random bits $b_i$, $c_i$. 
These signals are timed to arrive simultaneously at $T$.  
If $b_i = 0$, this signal instructs $T$ to 
carry out a measurement in the basis $b$ (
as in scheme IV) and report the result by a
classical broadcast in both directions (again, as in scheme V). 
If $b_i =1$, the signal instructs $T$ to send
the (unmeasured) qubit $\ket{\psi_i }$ in the
direction of $A_{c_i }$.  

\subsubsection{Informal discussion of scheme VI}

The following informal comments give some motivation for considering scheme VI,
but do not constitute a security proof.  

The aim of this design is to use the possibility that $b_i=1$ 
to prevent Eve from carrying out any form
of teleportation-like attack in which classical information
is extracted from the state $\psi_i$.  Such an operation
would imply that $\psi_i$ cannot be reliably reconstructed later,
which means that, if Eve performs it before she knows the value
of $b_i$, she risks detection if $b_i =1$. 
This ensures that Eve can only carry out teleportation-like
operations which (like standard teleportation) ensure that 
the ``teleported state'' takes the form $U \ket{\psi_i}$,
where $U$ is drawn from a finite list of possible unitary 
operations.  The list must include operations other than
the identity, since the density matrix of the ``teleported
state'', before reconstruction, is independent of $\ket{\psi_i}$.
However, there is no non-trivial unitary
operation which preserves all three bases $B_i$.   
This appears to leave Eve unable to 
carry out all the possible measurements required by the
protocol without reconstructing the state (which cannot
be done with the right timings, except by allowing the
state to arrive at the tag $T$).  

\subsection{Remarks on teleportation attacks}

Neither schemes IV and V share the vulnerability
of scheme III to the specific teleportation attacks described
above, since in both
cases there is {\it no} non-trivial unitary operation 
that leaves all the relevant bases (the three specified
bases in the case of scheme V, and the infinite set
of all possible bases in the case of scheme IV) invariant.

One might further hope that the schemes
are not vulnerable to general teleportation attacks, and more
generally that they are indeed secure against all
possible attacks.  

Since the general
set of operations that Eve might carry
out is rather large, it would certainly be desirable
if a security proof could be based on a specific
counter-physical implication of the form ``if Eve can spoof the tag, then it follows that they, perhaps
in collaboration with Alice, can implement some physical
operation known to be impossible''.  
An alternative proof strategy could be to identify sufficient constraints to
show that Eve's hands are effectively
tied. One would hope to show (at least for schemes IV and VI, possibly
also V) first that every possible operation
that Eve can carry out is provably detectable unless it is a
teleportation operation of a certain type, and
then that such teleportation attacks are also provably
detectable. (More precisely, one would like to prove both these
claims with some lower bound on the probability of detection per
spoofing attack.) 

We offer no security proof of either type here.  

\subsection{More general schemes}

Clearly, even in one dimension, the formulation of quantum tagging schemes allows
a plethora of options.   Alice could send both classical and quantum information
from both stations $A_0$ and $A_1$; the quantum information sent from $A_0$ and $A_1$
in any given round could be entangled, as could the quantum information used in
successive rounds; she could require any classical and quantum computation at $T$ that takes inputs
of the prescribed form and produces two (possibly entangled, possibly both classical
and quantum) output states to be returned to her sites.   

It would be very interesting to understand precisely which levels of security can be
attained by which types of tagging scheme, and how efficiently this can be done in
each case.   At present, to the best of our knowledge, these are open problems.   

\vskip15pt
\section{Brief History}
The possibility of quantum tagging protocols was first considered by one of us (AK) in 2002. 
The six protocols presented here, together with the teleportation attacks on the first
three, were variously invented and discussed by us during 2002-3.   A patent for a quantum tagging protocol
(which is not unconditionally secure, but appears unbreakable by present technology), 
based on notes filed for HP Labs Bristol in 2002, 
was granted and published in 2006.\cite{taggingpatent}    

Recently, other authors\cite{paperone, papertwo} have considered the possibility of quantum
tagging, and rediscovered some of the insecure protocols presented here, but apparently not the 
attacks on these protocols.   Refs. \cite{paperone,
papertwo} argue, incorrectly, that their protocols are in fact secure.
The protocol in Ref. \cite{papertwo} is a simpler version of Scheme III above, which we considered
in 2002.  It is breakable by the teleportation attack on scheme III described above.
The protocol in Ref. \cite{paperone} is a variation of a Bell state measurement scheme which we
also considered in 2002.  It is similarly breakable: Eve can intercept and store the two particles 
comprising quantum states
$ \ket{ \Gamma_i^{AB} }$ at sites equidistantly located either side of $T$, apply the unitaries
$( U_i^A )^{\dagger}$ and $( U_i^B )^{\dagger}$ to the respective states as soon as the
classical signals arrive at her sites, use teleportation to carry out a non-local Bell state measurement
on the resulting states, transmit the classical outcome data between her sites, calculate the measurement
result at both sites, and transmit the result to $A$ and $B$ so as to arrive at the expected times.

\section{Note added}

Some time after this work was circulated on the physics arxiv, 
papers developing further the 
results reported here were circulated
by Kent \cite{kentcryptotagging}, Lau and Lo\cite{laulo} and Buhrman et al.\cite{buhrmanetal}.
Ref. \cite{kentcryptotagging} shows that unconditionally secure quantum tagging is possible
in a scenario in which the tag is assumed to contain private data inaccessible to adversaries.  
Buhrman et al. show that schemes IV-VI, whose security was left
as an open question above, are insecure against eavesdroppers
with unbounded predistributed entanglement.  

\section{Acknowledgements}

AK was partially supported by an FQXi mini-grant and by Perimeter Institute for Theoretical
Physics. Research at Perimeter Institute is supported by the Government of Canada through Industry Canada and
by the Province of Ontario through the Ministry of Research and Innovation.
AK thanks Jonathan Oppenheim and Damian Pitalua-Garcia for helpful conversations.  
%
%

\end{document}